# Power Loss Modelling of GaN HEMT-based 3L-ANPC Three-Phase Inverter for different PWM Techniques

Salvatore Mita[1], Arjun Sujeeth[1], Giuseppe Aiello[1], Dario Patti[1], Francesco Gennaro[1], Giacomo Scelba[2], Mario Cacciato[2],

STMicroelectronics, Stradale Primo Sole 50, Catania, Italy [1]
University of Catania, Viale Andrea Doria 6, Catania, Italy [2]
E-Mail: giacomo.scelba@unict.it, francesco.gennaro@st.com

## Acknowledgements

This work was carried out within the ECSEL-JU project GaN4AP (GaN for Advanced Power Applications), under grant agreement no. 101007310. This Joint Undertaking receives support from the European Union's Horizon 2020 research and innovation programme.

## Abstract

The paper presents a straightforward modelling approach to compute the power loss distribution in GaN HEMT-based three-phase and three-level (3L) active neutral point clamped (ANPC) inverters, for different pulse width modulated techniques. Conduction and switching losses averaged over each PWM switching period are analytically computed by starting from the operating conditions of the AC load and data of GaN power devices. The accuracy of the proposed analytical approach is evaluated through a circuit-based power electronics simulation tool, applied to different carrier-based PWM strategies.

## Introduction

Nowadays, it is well recognized the key role of the electric traction as one of the prerequisites for the successful development of contemporary society, and the global electric vehicle market size is projected to a significant grow in the coming years even thanks to the policy objective set by many countries around the world of reducing greenhouse gas emissions from transport. Today, the rated power of the powertrain of electric vehicles generally varies from 60kW to 200kW, supplied by a lithium-ion battery pack whose capacity vary from 30kWh up to 90kWh, leading to optimistic estimated ranges lower than 450km. Although the recharging infrastructure is undergoing significant development in terms of charging points and installed power, most of the charging points are still limited to 50kW peak power, yielding to charging time not less than 45minutes ÷ 1h for reaching the 80% of full battery capacity when 50kW - 480V DC Fast Charging is used [1]-[3].

A shorter charge period could minimize the inconvenience of the driving range limitation from the consumer's point of view, and this goal could be achieved by using a higher charging power [1]-[5]. For instance, by increasing the fast-charging power level from 50 kW to 150 kW, the charging time is reduced by two-thirds. However, if the charging voltage level remains at the typical value of 400V, the current rating of the charging cable increases as well as the system power losses. Hence, some car makers are developing solutions to utilize an 800V DC bus beneficing in terms of shorter battery charge times [3]. This will however result in facing some technical issues related to the state of art technologies used in the electric powertrains, starting from the semiconductor technologies used to realize the electric traction inverter. Moreover, alternative power converter topologies could be considered to furtherly increase the efficiency, the reliability and the power density, while reduce the cost and weight of the power conversion processes.

Although still considered not suited for mostly motor drives, Gallium Nitride (GaN) power switches can potentially provide several benefits to the electric drives exploited in the traction, especially when high-speed electrical machines are suitably combined with GaN inverters topologies operating at high switching frequency [6]. In fact, GaN devices feature lower switching and conduction losses, higher power density, and higher-temperature operation compared to Si power switches. Moreover, GaN devices possess much less parasitic components and lower on-state resistance which make it more suitable than the SiC device in hard switched applications. Hence, this technology can furtherly contribute to realize electric drives for traction featuring extreme compactness, high efficiencies, robustness and reduced weight, all factors contributing to further extend the range of the vehicle. Moreover, the increase of DC bus voltage architectures (800V) in the next generation of EVs developed with the goal of enabling faster charging rates, will have direct consequence to increase the losses of the standard 2L VSIs based on Si and SiC devices, thus reducing the system efficiency. This makes the use of GaN multilevel inverters an attractive solution for designing very compact and high-efficiency traction inverters overcoming the breakdown voltage limits (650V) of the actual GaN technologies. Moreover, thanks to the reduced output voltage steps, the use of multilevel converters is an effective solution to reduce the voltage stress and thus the electrical aging in electrical traction machines. Several configurations of multilevel topologies have been presented in the technical literature, among them the neutral point clamped (NPC), the Active NPC (ANPC), and T-type NPC (TNPC) [7]-[12]. The 3L-ANPC inverter represents a good compromise between performances, compactness, efficiency, and investment cost and is considered one of the promising multilevel converter topologies for medium voltage application, even for electric traction application [10]-[12]. The 3L-ANPC topology is shown in Fig. 1. It consists of six switches for each leg, which allow to generate three voltage levels. In past literatures several modulation strategies have been presented with the main goal of minimizing conduction and switching power loss, but in most of the cases, sophisticated mathematical or circuital-based solutions have been used to carry out the loss distribution and they were specific only for a single modulation strategy [13]-[18].

This paper presents an accurate analytical modelling devoted to the power losses estimation for GaN HEMT-based three-phase 3L-ANPC inverters for different modulation techniques. The main goal of the proposed investigation is to evaluate, among a certain number of well-known PWM modulation strategies [13], [15], the power loss that allows to get a modulation technique with minimum power losses and better thermal distribution. The paper is structured to first provide the theoretical description of the ANPC operation, and the mathematical model used to compute the power loss. Then, the presentation of different modulation techniques is detailed provided, underlying pros and cons. Finally, some experimental tests are also provided to validate the accuracy of the theoretical modelling. Although the methodology presented in this paper has been applied to some modulation methods, it could be straightforwardly extended even to other PWM modulation strategies.

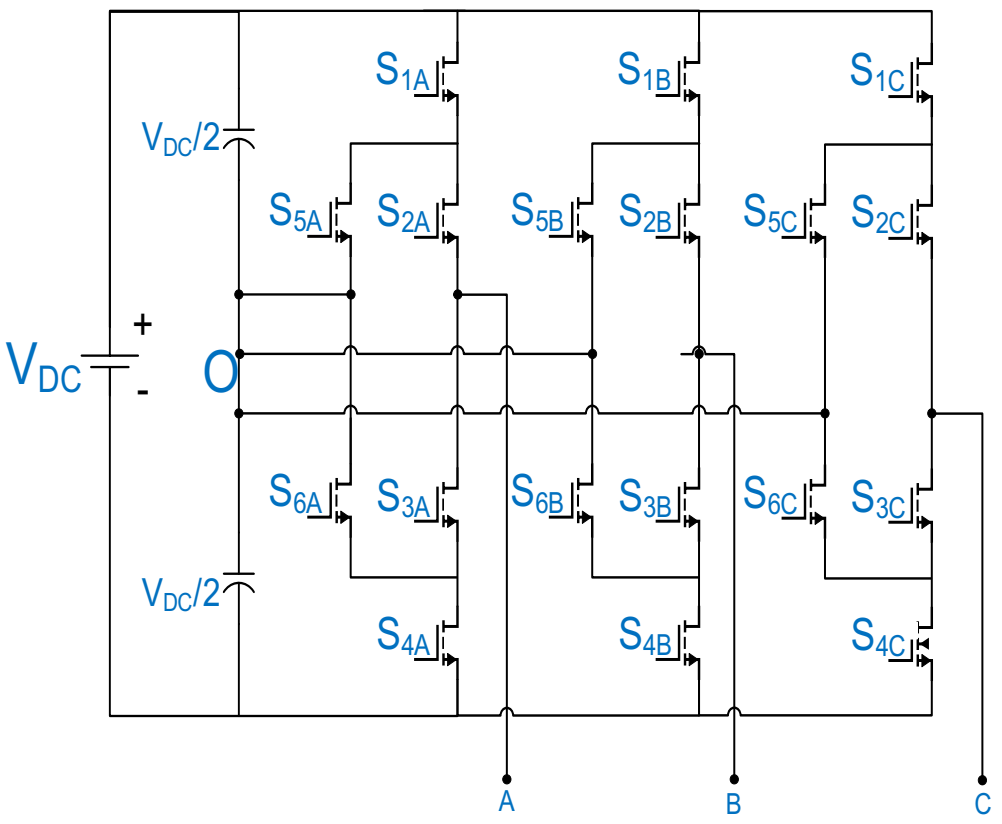

(a)

| Table I: | |
|---|---|
| **Parameter** | **Value** |
| DC Bus Voltage [$V_{DC}$] | 800V |
| Line-to-Line Voltage [$V_{LL}$] | 400V |
| cos $\varphi$ [$PF$] | 0.9 |
| Root Mean Square Current [$I_{RMS}$] | 6.95A |
| Active Power [$P$] | 4.5kW |
| Resistance Load [$R_L$] | 25.65Ω |
| Inductance Load [$L_L$] | 39.5mH |
| Fundamental Frequency [$f_e$] | 50Hz |
| Switching Frequency [$f_{sw}$] | 50kHz |
| Modulation Index [$m$] | 0.7 |
| Resistance Drain-Source [$R_{DS}$] | 120mΩ |
| Junction Temperature [$T_j$] | 80°C |

(b)

Fig. 1: (a) Three-phase active neutral point clamped (ANPC) GaN-based topology (b) main power converter and load settings.

## Three-Phase 3L-Active Neutral Point Clamped Inverter

Each leg of the 3L ANPC inverter consists of six bidirectional-conducting-unidirectional-blocking switches [4], which allow to generate three voltage levels. This converter can be considered an evolution of the standard neutral point clamped (NPC) inverter. The last name comes from the two diodes connected in anti-parallel that are used to "clamp" the output voltage to the neutral point of the DC circuit when the zero-voltage level is required. The output current direction determines whether the neutral point current flows through the upper or the lower current path.
Compared to the NPC, two additional switches $S_5$ and $S_6$ substitute the clamping diodes in the neutral point connection, allowing to actively clamp the output to the neutral point of the DC circuit. In this way, when the ANPC inverter outputs zero state, there are multiple redundant loops to choose from. By rationally selecting the zero-state loop, the loss balance of each device can be achieved [15]. Hereafter, some specific PWM strategies for the 3L ANPC inverter are described.

### Carrier based Modulation Techniques

In the following analysis different PWM modulation techniques are considered [13], [15]: (a) DNPC modulation, (b) ANPC modulation with same-side clamping. (c) ANPC modulation with opposite-side clamping. (d) ANPC modulation with full-path clamping; these strategies can be applied to both sinusoidal and third harmonic injection SPWM.

- *DNPC Modulation (DNPC)*

The DNPC modulation keeps the two clamping switches $S_5$ and $S_6$ constantly OFF in the entire fundamental period, similarly to the diode clamped NPC. On the contrary, the gate signals for $S_1 - S_6$ are generated according to the phase disposition PWM, as shown in Fig. 2a. Note that $S_1$ is switched in complementary with $S_3$, and $S_2$ is complementary to $S_4$. The command signal applied to $S_1$ is generated by comparing the reference sinusoidal $V_{ref}$ to $V_{tri1}$, whereas the switching pattern for $S_4$ is generated by comparing $V_{ref}$ to $V_{tri2}$. The switches $S_1 - S_4$ are pulse width modulated at the switching frequency $f_{sw}$ in half fundamental cycle while keep a constant state in the other half period.

- *ANPC Modulation With Same-Side Clamping (ANPC-SSCM)*

This PWM implementation forces the path of the load current to flow through the upper cell during the positive half cycle and vice-versa in the negative half; as shown in Fig. 2b, three pairs of complementary switches are used: $S_1$ and $S_5$, $S_2$ and $S_3$, $S_4$ and $S_6$. The command signal of $S_1$ is generated by comparing $V_{ref}$ to $V_{tri1}$ while $S_4$ is commutated by comparing $V_{ref}$ to $V_{tri2}$ and $S_2$ commutates for every half cycle by comparing $V_{ref}$ to $0$.
The switches $S_1$, $S_4$, $S_5$ and $S_6$ are modulated at $f_{sw}$ in half of the fundamental period while remains at a constant state in the other half period. The inner switches $S_2$ and $S_3$ are commutated at the fundamental frequency $f_e$.

- *ANPC Modulation With Opposite-Side Clamping (ANPC-OSCM)*

An alternative PWM implementation can be achieved by using the lower neutral path for the top cell commutation and the upper neutral path for the bottom cell commutation, which is called "opposite-side clamping". As shown in Fig. 2c, the complementary switch pairs are the same of *ANPC-SSCM*, but the switching patterns are quite different. The command signals of $S_1$ and $S_6$ is carried out by comparing $V_{ref}$ to $0$ whereas $S_4$ and $S_5$ are the complementary. The command signal of $S_2$ is generated by comparing $V_{ref}$ to $V_{tri1}$ when $V_{ref}$ is greater than 0 while $V_{ref}$ is compared to $V_{tri2}$ when it is less than 0. Only the inner switches $S_2$ and $S_3$ are modulated at $f_{sw}$, while the others are all switched at $f_e$.

- *ANPC Modulation With Full-Path Clamping: (ANPC-FPCM)*

Instead of exploiting only one current path during the neutral states, the upper path and the lower path can be used together; this strategy is referred in the literature as "full-path clamping". The two paralleled current paths can reduce the on-state resistance and thus the conduction loss during the neutral states. As shown in Fig. 2d, the devices $S_3$ and $S_5$ are simultaneously turned on and off and complementary to $S_1$, while the devices $S_4$ and $S_6$ simultaneously turned on and off and the command signals are complementary to $S_2$. This modulation is achieved by comparing $V_{ref}$ to $V_{tri1}$ for driving $S_1$ and $V_{ref}$ to $V_{tri2}$ for driving $S_4$ and $S_6$.

The above modulation methods can be also extended to the case of third harmonic injection PWM, where a third harmonic zero sequence voltage waveform set is added to the reference sinusoidal voltages [13]. The last approach allows to increase the DC bus voltage utilization of 15%.

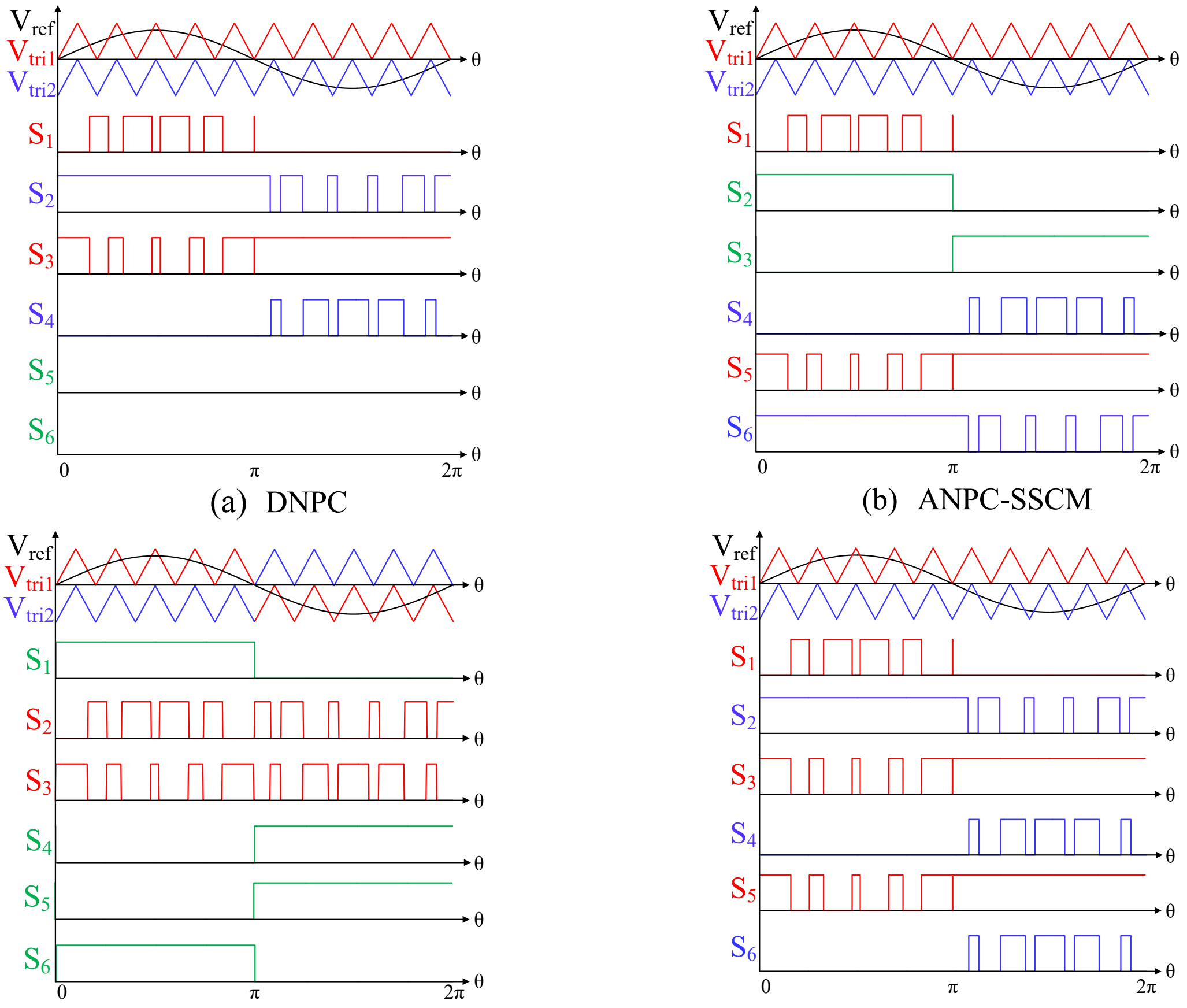

(a) DNPC (b) ANPC-SSCM (c) ANPC-OSCM (d) ANPC-FPCM

Fig. 2: Modulation strategies and corresponding switching patterns.

# Power Loss Computation

Hereafter, the analytical approach to determine the conduction and switching losses of the devices composing the 3L-ANPC inverter is presented for the considered sinusoidal PWM carrier-based modulation strategies.

## Conduction Losses

For each GaN power device the conduction losses are computed according to (1), where the RMS current is given by (3); we are assuming a sinusoidal load current (2) and the duty cycle $d(\theta)$ of one phase of the PWM voltage waveform is defined as a quantity variable with the modulation index $m$ and power factor $cos(\varphi)$, [19]. The load current is identified by the peak value $I_p$ and phase angle $\varphi$, while the angular position $\theta$ is given by $\omega_e t$, where $\omega_e$ is the fundamental voltage harmonic angular frequency.

$$P_{cond} = I_{RMS}^2 R_{DS(ON)} \quad (1) \qquad I(\theta) = I_p \sin(\theta - \varphi) \quad (2) \qquad I_{RMS} = \sqrt{\frac{1}{\pi}\int_0^{2\pi}(I_p^2 \sin^2(\theta-\varphi))d(\theta)} \quad (3)$$

The power device currents for the four type of sinusoidal carrier based modulation techniques considered in this study are shown in Fig. 2, with the power converter operating under the conditions indicated in Tab. I. The carrier-based modulation techniques feature a symmetrical behavior: $I_{rms(S1)}=I_{rms(S4)}$, $I_{rms(S2)}=I_{rms(S3)}$, $I_{rms(S5)}=I_{rms(S6)}$, thus the conduction losses computation can be limited to the top half GaN devices of the power converter i.e. $S_1$, $S_2$ and $S_5$. The computation of the RMS currents $I_{RMS}$ (3) for each device are reported in Table. II. The computation has been carried out by considering the reverse conduction mechanism of GaN HEMT, thus considering the two current parallel paths for each device.

The state of the generic power switch $S_j$ associated to forward and reverse current conductions is indicated with $S_{jF}$ and $S_{jR}$, respectively. Same approach can be extended even to other modulation strategies, such as the third harmonic injection PWM techniques.

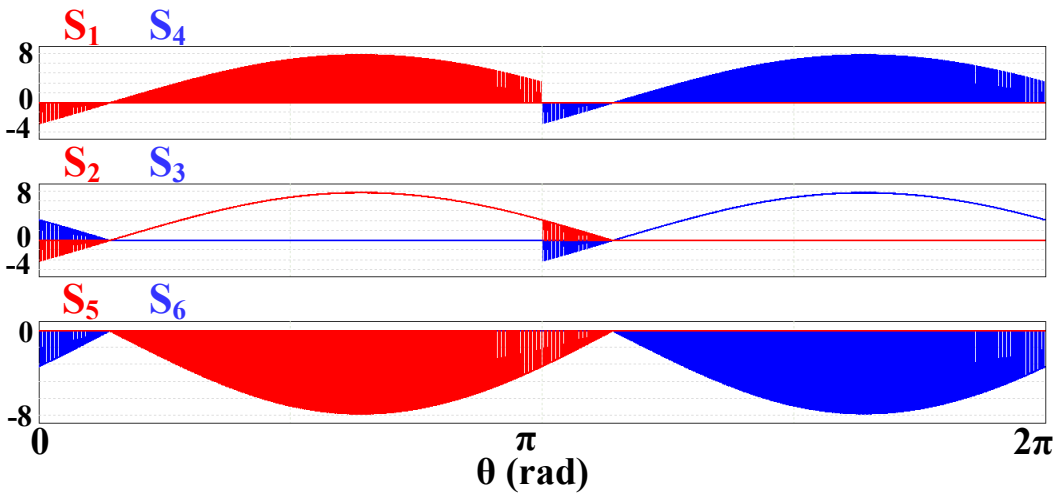

(a) DNPC.

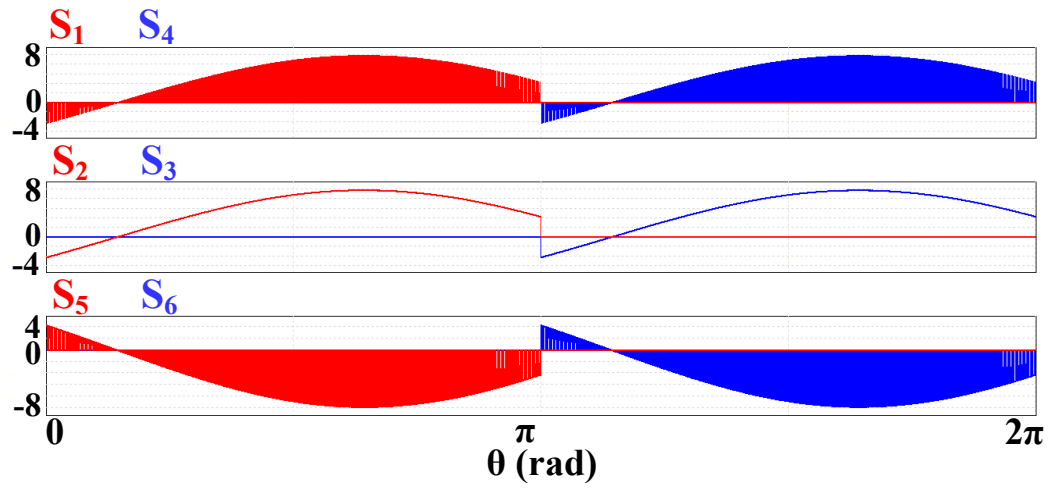

(b) ANPC-SSCM.

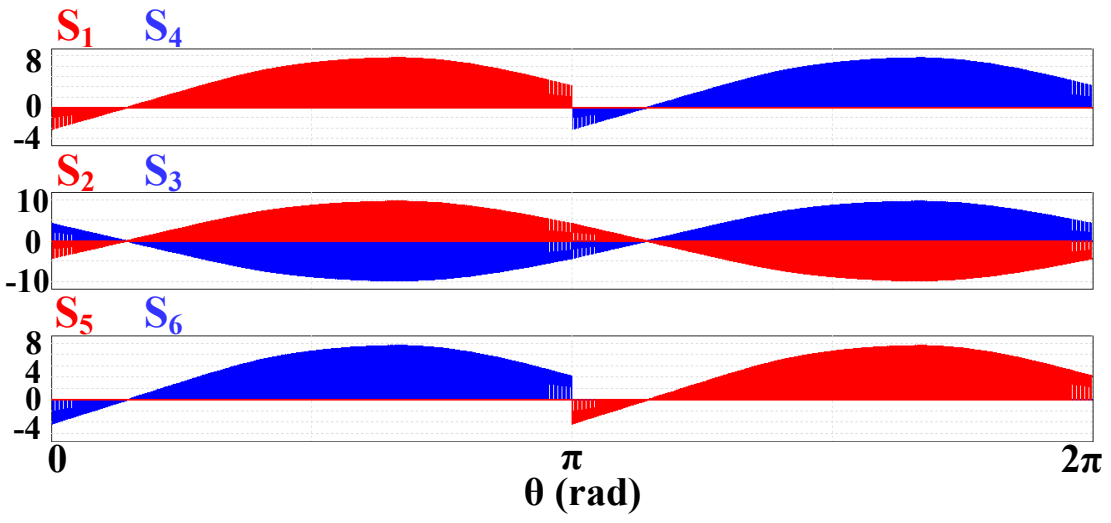

(c) ANPC-OSCM.

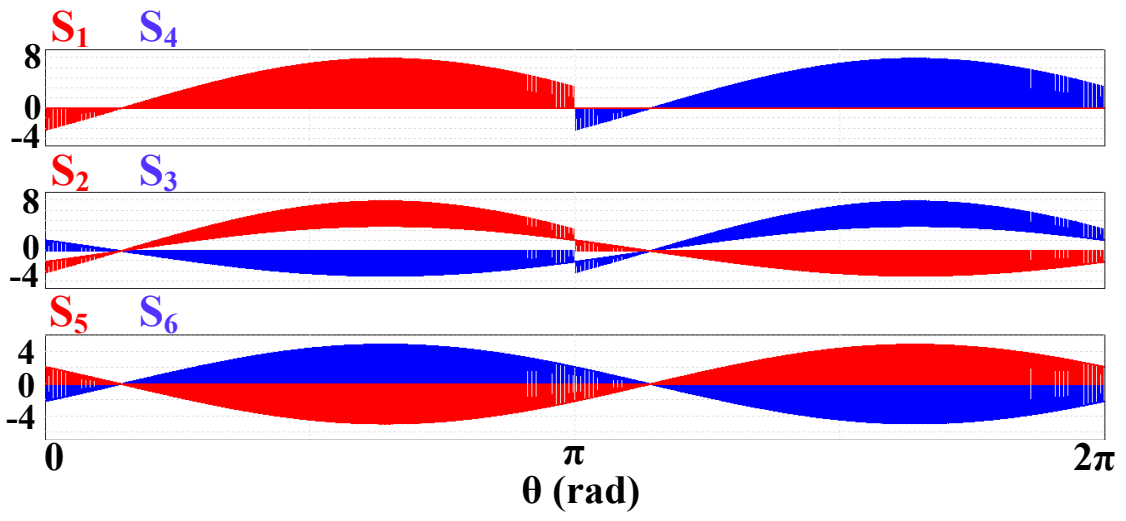

(d) ANPC-FPCM.

Fig. 2: Currents waveforms flowing through the GaN power devices in the ANPC for sinusoidal PWM modulation techniques (cos φ = 0.9).

## Switching Losses

An analytical approach for the estimation of switching losses is provided below. The input data required for switching losses estimation are given by the characteristic curves of the GaN energy losses ($E_{ON}$ and $E_{OFF}$), expressed as a function of the current magnitude $I_{ds}$ and gate resistor $R_g$ (T=25°C), as shown in Fig. 3.

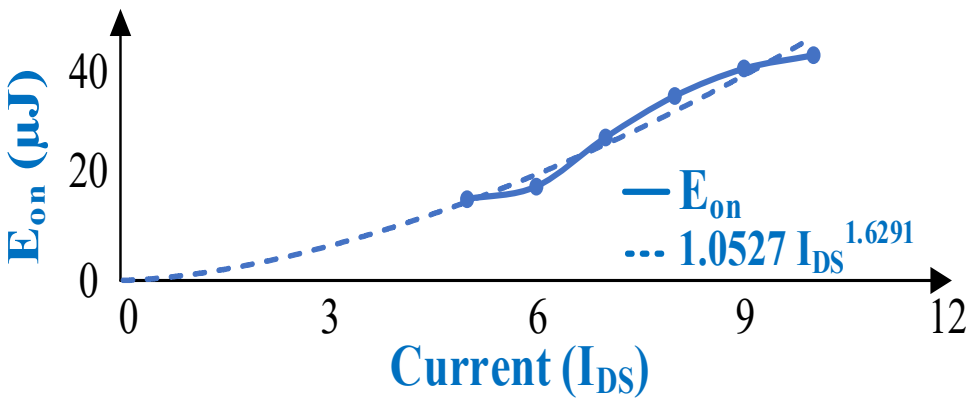

*(a)*

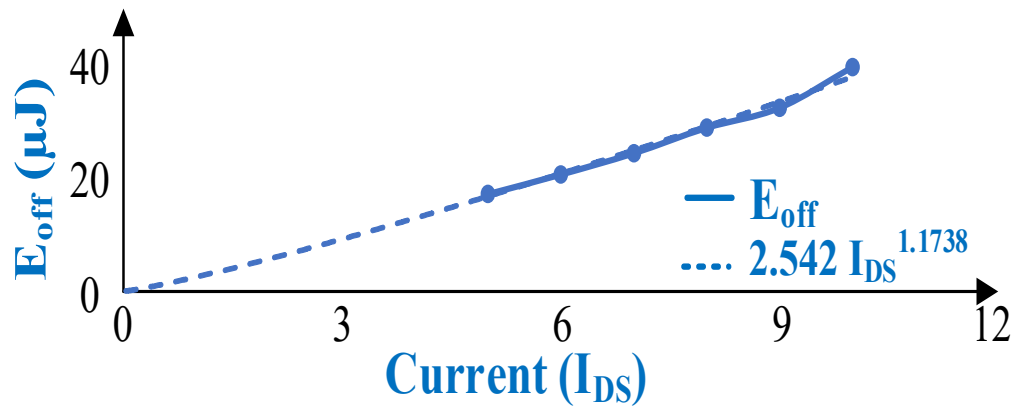

*(b)*

Fig. 3: Turn ON (a) and turn OFF (b) switching energy losses of the GaN HEMT under test.

It has been experienced that energy losses curves can be easily approximated by sampling and interpolating with a second order equation. For the considered device, the expression used to calculate the average value of the energy losses $E_{ON}$ and $E_{OFF}$ and powers $P_{ON}$ and $P_{OFF}$ for the generic GaN HEMT device $S_j$ in each switching period are given by (4)-(7).

$$E_{ON} = (1.0527 * |I_{DS}|^{1.6291}) * 10^{-6} \text{ [J]} \quad (4) \qquad E_{OFF} = (2.542 * |I_{DS}|^{1.1738}) * 10^{-6} \text{ [J]} \quad (5)$$

$$P_{ON} = (1.0527 * |I_{DS}|^{1.6291}) * 10^{-6} * f_{sw} \text{ [W]} \quad (6) \qquad P_{OFF} = (2.542 * |I_{DS}|^{1.1738}) 10^{-6} * f_{sw} \text{ [W]} \quad (7)$$

By considering the number of commutations $n$ of each power switch within the fundamental period, the average switching loss are given by:

$$P_{sw} = \frac{1}{n} \sum_{i=1}^{n} (P_{ONi} + P_{OFFi}) \qquad (8)$$

## Table II: RMS currents for the considered sinusoidal modulation strategies

| PWM strategy | $S_j$ | $d(\theta)$ ; interval | $I_{rms}$ |
|---|---|---|---|
| DNPC (a) | $S_{1F}$ | $m\sin(\theta)$ ; $\varphi<\theta<\pi$ | $\frac{I_P}{\sqrt{2\pi}}\sqrt{\frac{m(\cos(\varphi)+1)^2}{3}}$ |
| | $S_{1R}$ | $m\sin(\theta)$ ; $0<\theta<\varphi$ | $\frac{I_P}{\sqrt{2\pi}}\sqrt{\frac{m(\cos(\varphi)-1)^2}{3}}$ |
| | $S_{2F}$ | $1$ ; $\varphi<\theta<\pi$<br>$1+m\sin(\theta)$; $\pi<\theta<\pi+\varphi$ | $\frac{I_P}{\sqrt{2\pi}}\sqrt{\frac{\pi}{2}-\frac{(m\cos(\varphi)-1)^2}{3}}$ |
| | $S_{2R}$ | $m\sin(\theta)$ | $\frac{I_P}{\sqrt{2\pi}}\sqrt{\frac{m(\cos(\varphi)-1)^2}{3}}$ |
| | $S_{5R}$ | $1-m\sin(\theta)$ ; $\varphi<\theta<\pi$<br>$1+m\sin(\theta)$ ; $\pi<\theta<\pi+\varphi$ | $\frac{I_P}{\sqrt{2\pi}}\sqrt{\frac{\pi}{2}-\frac{2m}{3}(1+\cos(\varphi)^2)}$ |
| ANPC-SSCM (b) | $S_{1F}$ | $m\sin(\theta)$ ; $\varphi<\theta<\pi$ | $\frac{I_P}{\sqrt{2\pi}}\sqrt{\frac{m(\cos(\varphi)+1)^2}{3}}$ |
| | $S_{1R}$ | $m\sin(\theta)$ ; $0<\theta<\varphi$ | $\frac{I_P}{\sqrt{2\pi}}\sqrt{\frac{m(\cos(\varphi)-1)^2}{3}}$ |
| | $S_{2F}$ | $1$ ; $\varphi<\theta<\pi$ | $\frac{I_P}{\sqrt{2\pi}}\sqrt{\frac{\pi}{2}-\frac{\varphi}{2}-\frac{\sin(2\varphi)}{4}}$ |
| | $S_{2R}$ | $1$ ; $0<\theta<\varphi$ | $\frac{I_P}{\sqrt{2\pi}}\sqrt{\frac{\pi}{2}-\frac{\sin(2\varphi)}{4}}$ |
| | $S_{5F}$ | $1-m\sin(\theta)$ ; $0<\theta<\varphi$ | $\frac{I_P}{\sqrt{2\pi}}\sqrt{\frac{\varphi}{2}-\frac{\sin(2\varphi)}{4}-\frac{m(\cos(\varphi)-1)^2}{3}}$ |
| | $S_{5R}$ | $1-m\sin(\theta)$ ; $\varphi<\theta<\pi$ | $\frac{I_P}{\sqrt{2\pi}}\sqrt{\frac{\pi}{2}-\frac{\varphi}{2}-\frac{\sin(2\varphi)}{4}-\frac{m(\cos(\varphi)-1)^2}{3}}$ |
| ANPC-OSCM (c) | $S_{1F}$ | $m\sin(\theta)$ ; $\varphi<\theta<\pi$ | $\frac{I_P}{\sqrt{2\pi}}\sqrt{\frac{m(\cos(\varphi)+1)^2}{3}}$ |
| | $S_{1R}$ | $m\sin(\theta)$ ; $0<\theta<\varphi$ | $\frac{I_P}{\sqrt{2\pi}}\sqrt{\frac{m(\cos(\varphi)-1)^2}{3}}$ |
| | $S_{2F}$ | $m\sin(\theta)$; $\varphi<\theta<\pi+\varphi$ | $\frac{I_P}{\sqrt{2\pi}}\sqrt{\frac{4m\cos\varphi}{3}}$ |
| | $S_{2R}$ | $m\sin(\theta+\pi)$ ; $0<\theta<\varphi$<br>$1-m\sin(\theta+\pi)$; $\pi+\varphi<\theta<2\pi$ | $\frac{I_P}{\sqrt{2\pi}}\sqrt{\frac{\pi}{2}-\frac{\varphi}{2}+\frac{\sin(2\varphi)}{4}-\frac{m(\cos(\varphi)-1)^2}{3}-\frac{m(\cos(\varphi)+1)^2}{3}}$ |
| | $S_{5F}$ | $1-m\sin(\theta)$; $0<\theta<\varphi$ | $\frac{I_P}{\sqrt{2\pi}}\sqrt{\frac{\pi}{2}-\frac{\varphi}{2}+\frac{\sin(2\varphi)}{4}-\frac{m(\cos(\varphi)+1)^2}{3}}$ |
| | $S_{5R}$ | $1-m\sin(\theta)$; $\varphi<\theta<\pi$ | $\frac{I_P}{\sqrt{2\pi}}\sqrt{\frac{\varphi}{2}-\frac{\sin(2\varphi)}{4}-\frac{m(\cos(\varphi)-1)^2}{3}}$ |
| ANPC-FPCM (d) | $S_{1F}$ | $m\sin(\theta)$ ; $\varphi<\theta<\pi$ | $\frac{I_P}{\sqrt{2\pi}}\sqrt{\frac{m(\cos(\varphi)+1)^2}{3}}$ |
| | $S_{1R}$ | $m\sin(\theta)$ ; $0<\theta<\varphi$ | $\frac{I_P}{\sqrt{2\pi}}\sqrt{\frac{m(\cos(\varphi)-1)^2}{3}}$ |
| | $S_{2F}$ | $m\sin(\theta)$; $\varphi<\theta<\varphi+\pi$<br>$1-m\sin(\theta)$; $\varphi<\theta<\varphi+\pi$ | $\frac{I_P}{\sqrt{2\pi}}\sqrt{\frac{\pi}{8}+(m\cos\varphi)}$ |
| | $S_{2R}$ | $1-2m\sin(\theta+\pi)$; $0<\theta<\varphi$<br>$1-m\sin(\theta+\pi)$; $\varphi<\theta<\varphi+\pi$ | $\frac{I_P}{\sqrt{2\pi}}\sqrt{\frac{\pi}{8}-\frac{m(\cos(\varphi)+1)^2}{12}}$ |
| | $S_{5F}$ | $1-m\sin(\theta+\pi)$;<br>$\pi+\varphi<\theta<\varphi+2\pi$ | $\frac{I_P}{\sqrt{2\pi}}\sqrt{\frac{\pi}{8}-\frac{m\cos(\varphi)}{3}}$ |
| | $S_{5R}$ | $1-m\sin(\theta)$; $\varphi<\theta<\varphi+\pi$ | $\frac{I_P}{\sqrt{2\pi}}\sqrt{\frac{\pi}{8}-\frac{m\cos(\varphi)}{3}}$ |

## Study Case

The proposed modelling has been tested under different operating conditions, and the following results are referred to the operating conditions listed in Tab. I. The budget of conduction and switching losses associated to each power switch for the considered PWM techniques are shown in Fig. 4, while the total losses for each device are displayed in Fig. 5. Note an unequal loss distribution among the devices depending on the modulation strategy, which is less evident in case of ANPC-FPCM.

The effectiveness and accuracy of the proposed approach has been initially evaluated by realizing the same power conversion system with PSIM, a circuit-based power electronics simulator containing a specific tool allowing to include the electrical and thermal characterization curves in the power device models. The circuital scheme and a picture of the device database editor are displayed in Fig. 7 and Fig. 8. Current and voltage acquisitions have been performed to compute the loss and delivered power. The error between the results obtained with the analytical approach and the circuit-based simulator are displayed in Fig. 6, confirming the feasibility of the method with satisfying achievements, i.e. differences lower than 3%.

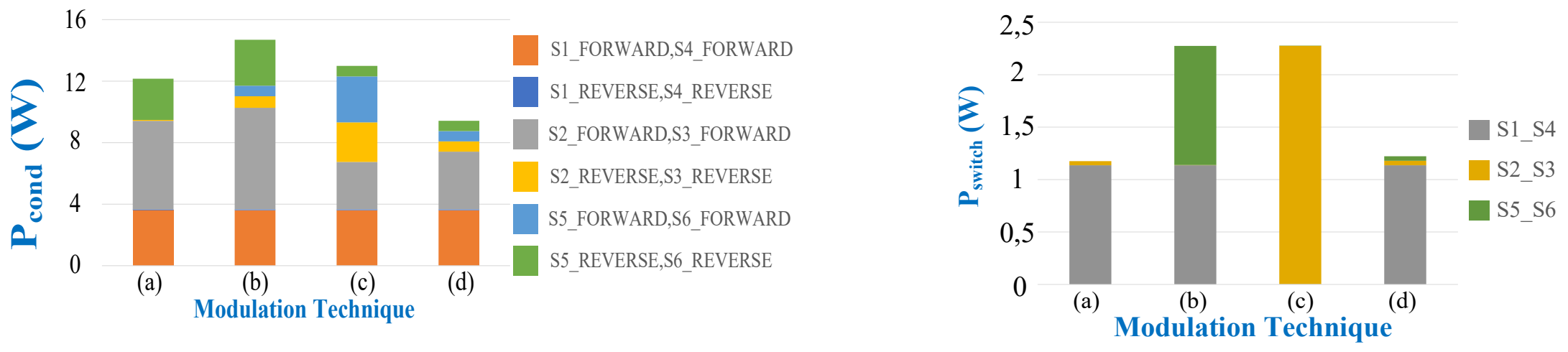

(a) (b)

Fig. 4: (a) Conduction Losses and (b) Switching losses for the sinusoidal PWM techniques.

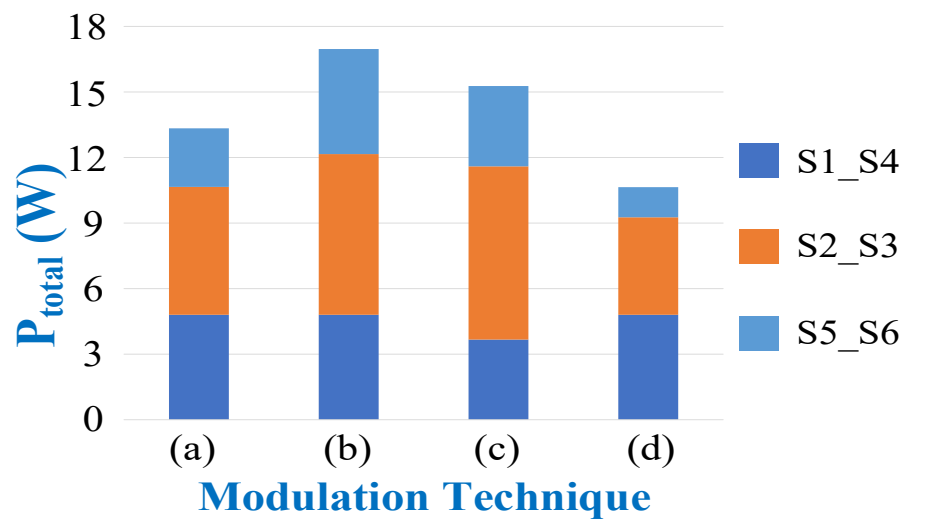

Fig. 5: Total losses distribution in the ANPC inverter leg for each modulation techniques.

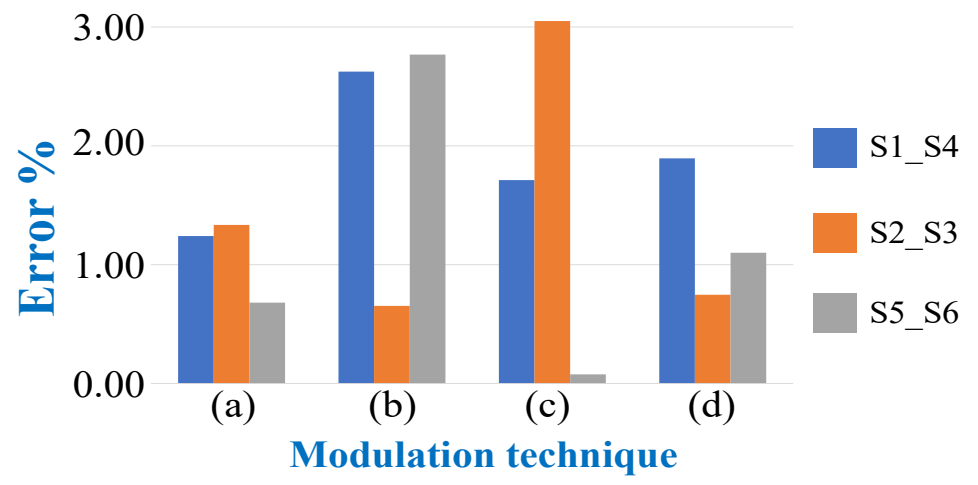

Fig. 6: Difference between total losses computed according to the proposed approach and simulations.

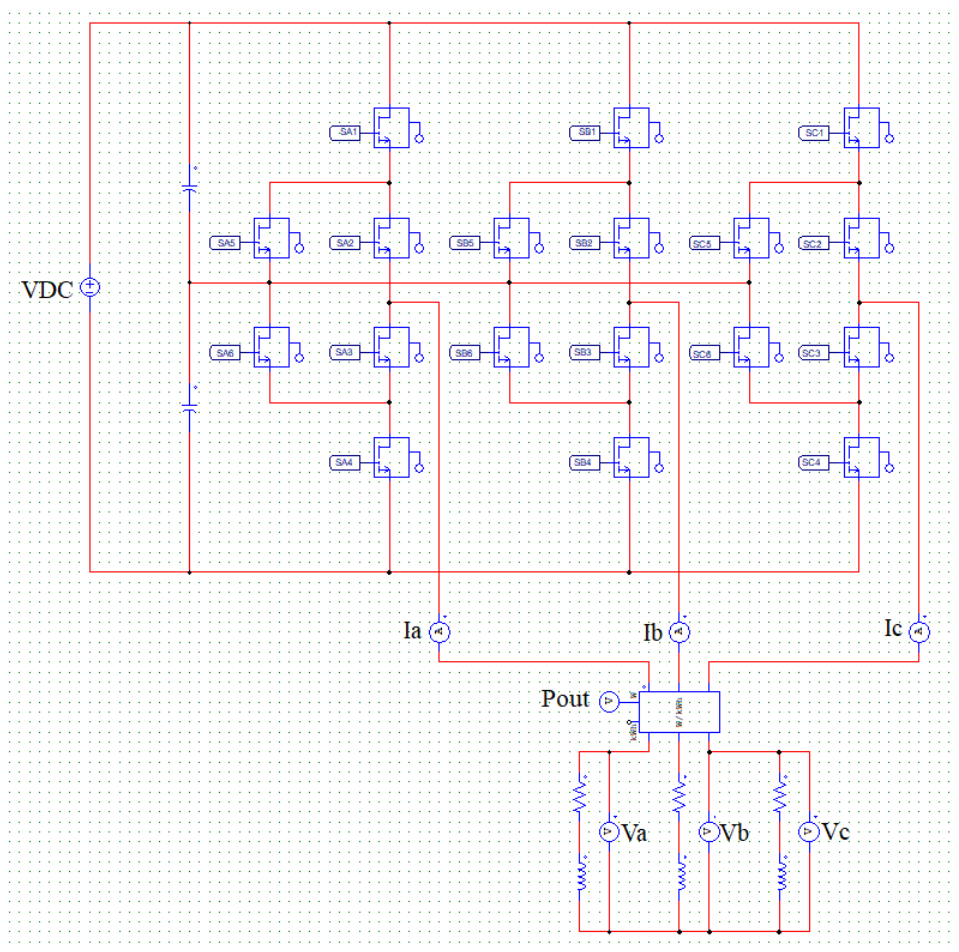

Fig. 7: Circuital modelling of the ANPC realized in PSIM for losses investigation.

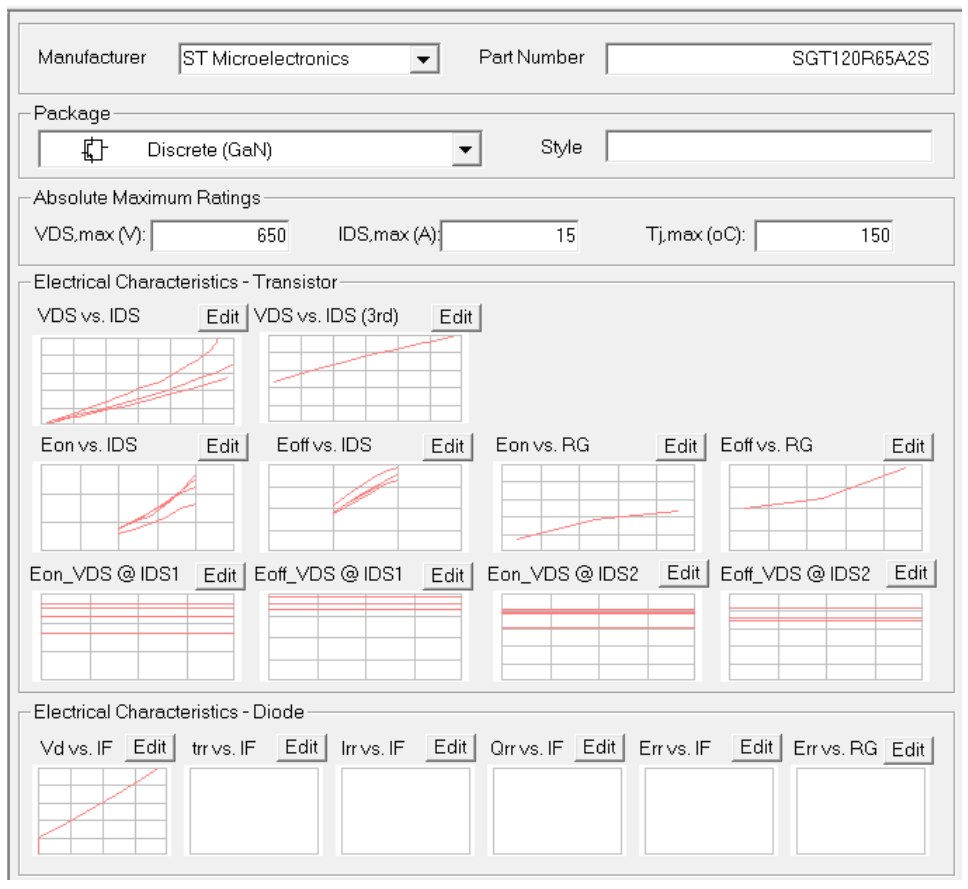

Fig. 8: Device database editor used to set the electrical and thermal characteristics of the GaN devices.

## Experimental Results

An experimental test bench has been arranged to carry out some preliminary results to be compared with that coming from the modelling. The single phase of an ANPC was realized by suitably connecting three half bridge converters, including GaN devices whose specifications are listed in Table III. The modulation strategies have been implemented in a STM32-G474RE control board, based on the high-performance Arm® Cortex®-M4 32-bit RISC core. The switching frequency of GaN devices is $f_{sw}$=50kHz, while the dead time is 100ns. The DC bus voltage has been set to 200V, while the inverter output has been connected to an ohmic-inductive load given by a series connection of $R=20.65\Omega$ and $L=1mH$. Fig. 9 displays the test rig, while Fig. 10 shows a zoomed in view of the single phase ANPC.

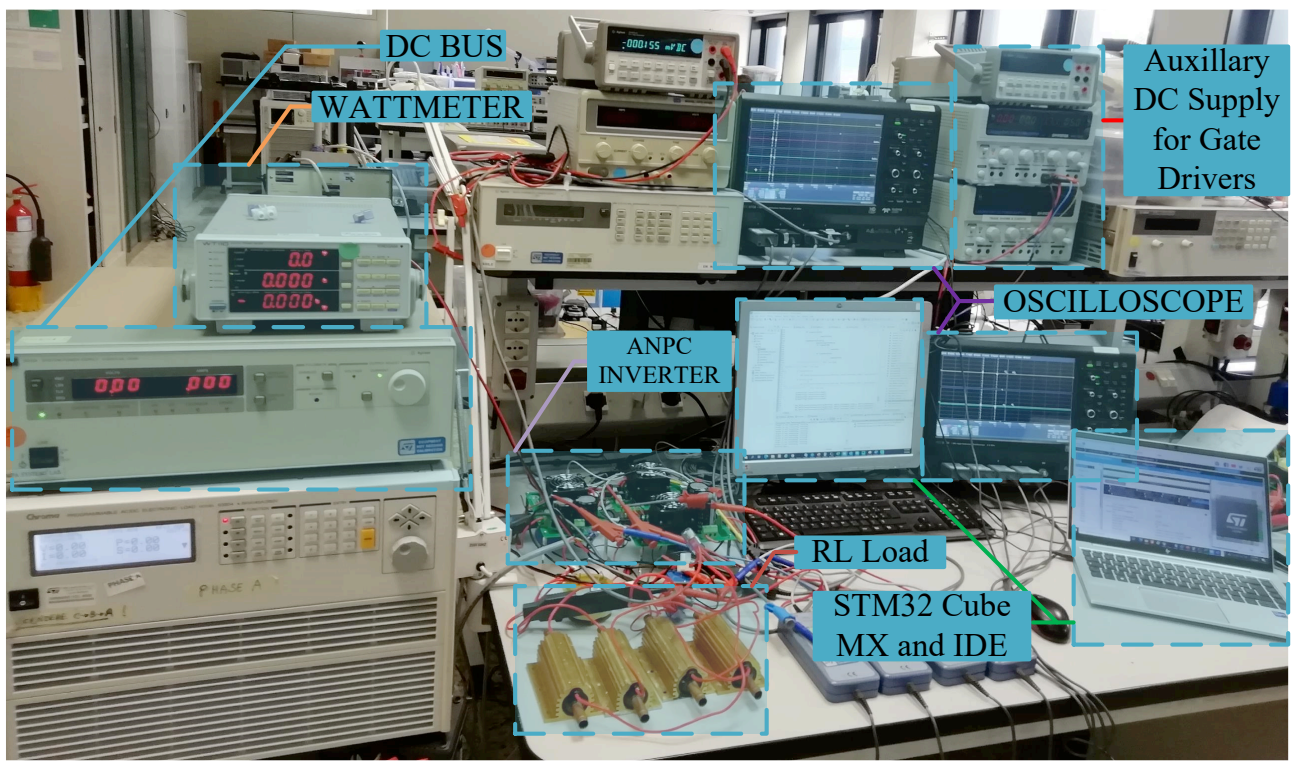

Fig. 9: Test Bench.

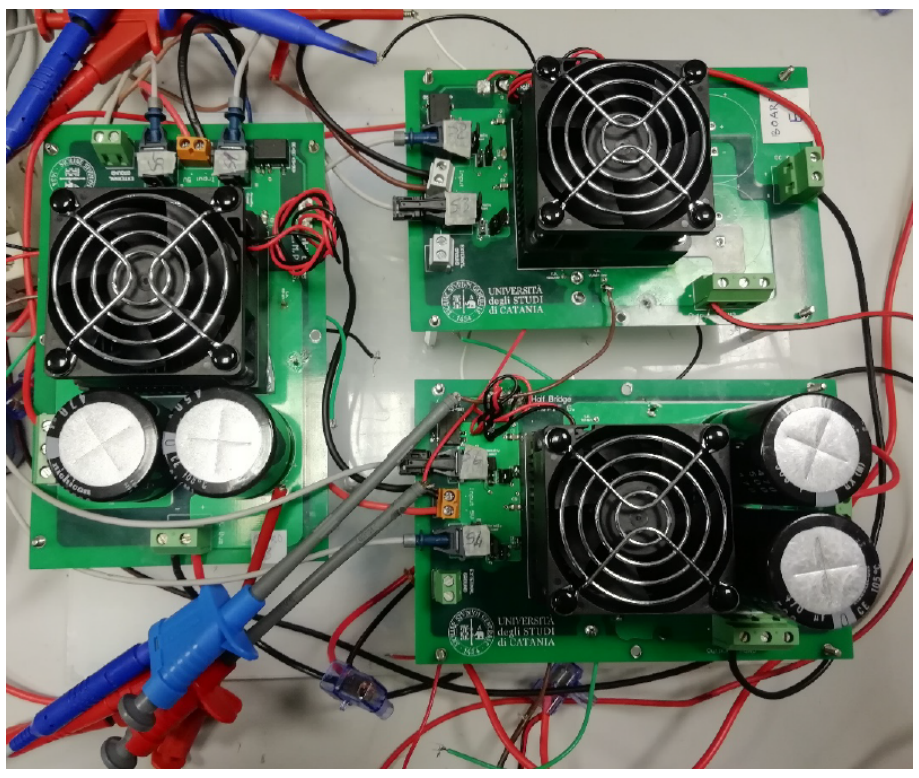

Fig. 10 GaN ANPC.

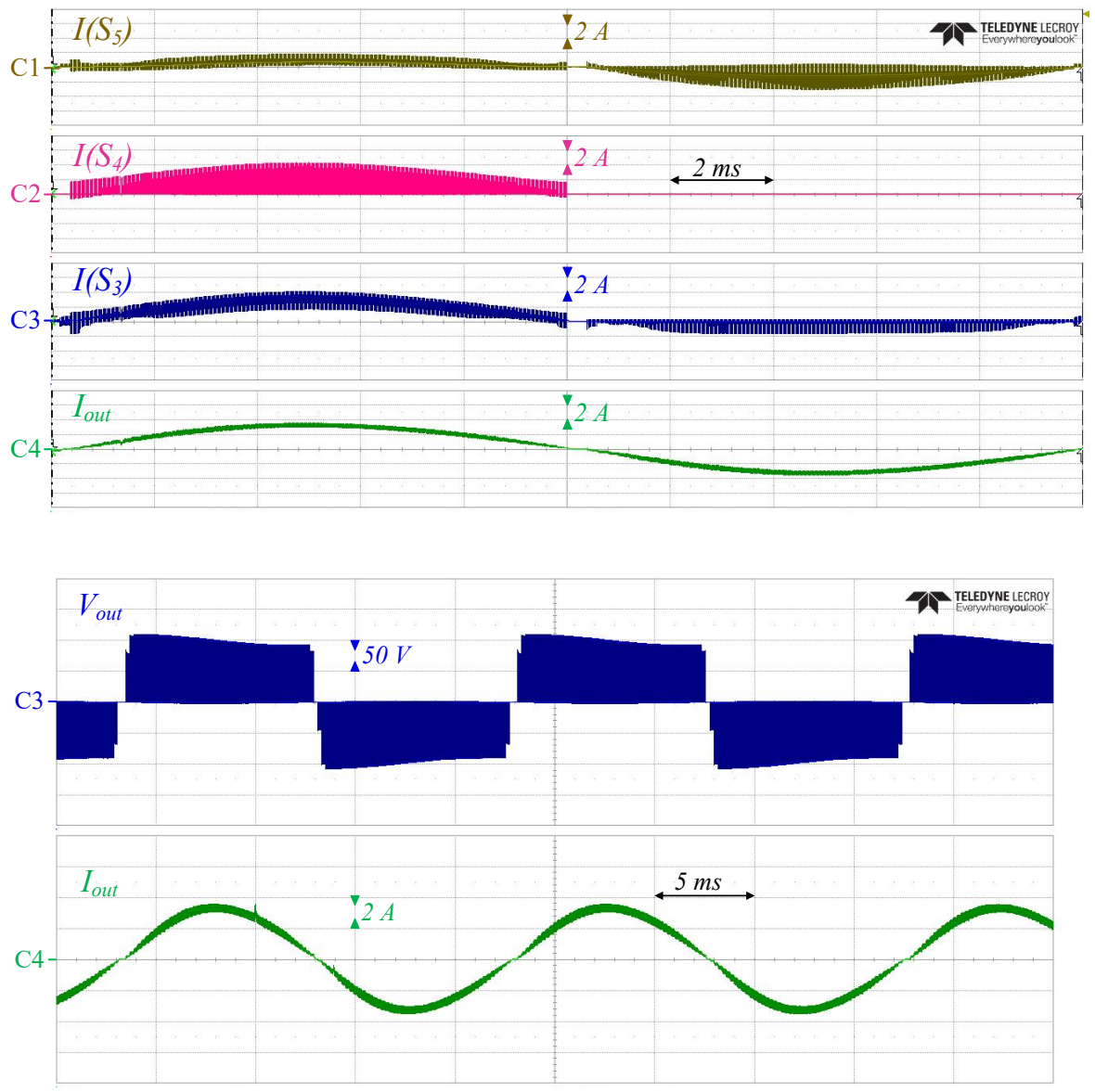

Fig. 11: Implementation of ANPC-SSCM: $V_{in}$=200V $I_{out}$=3A m=0.7.

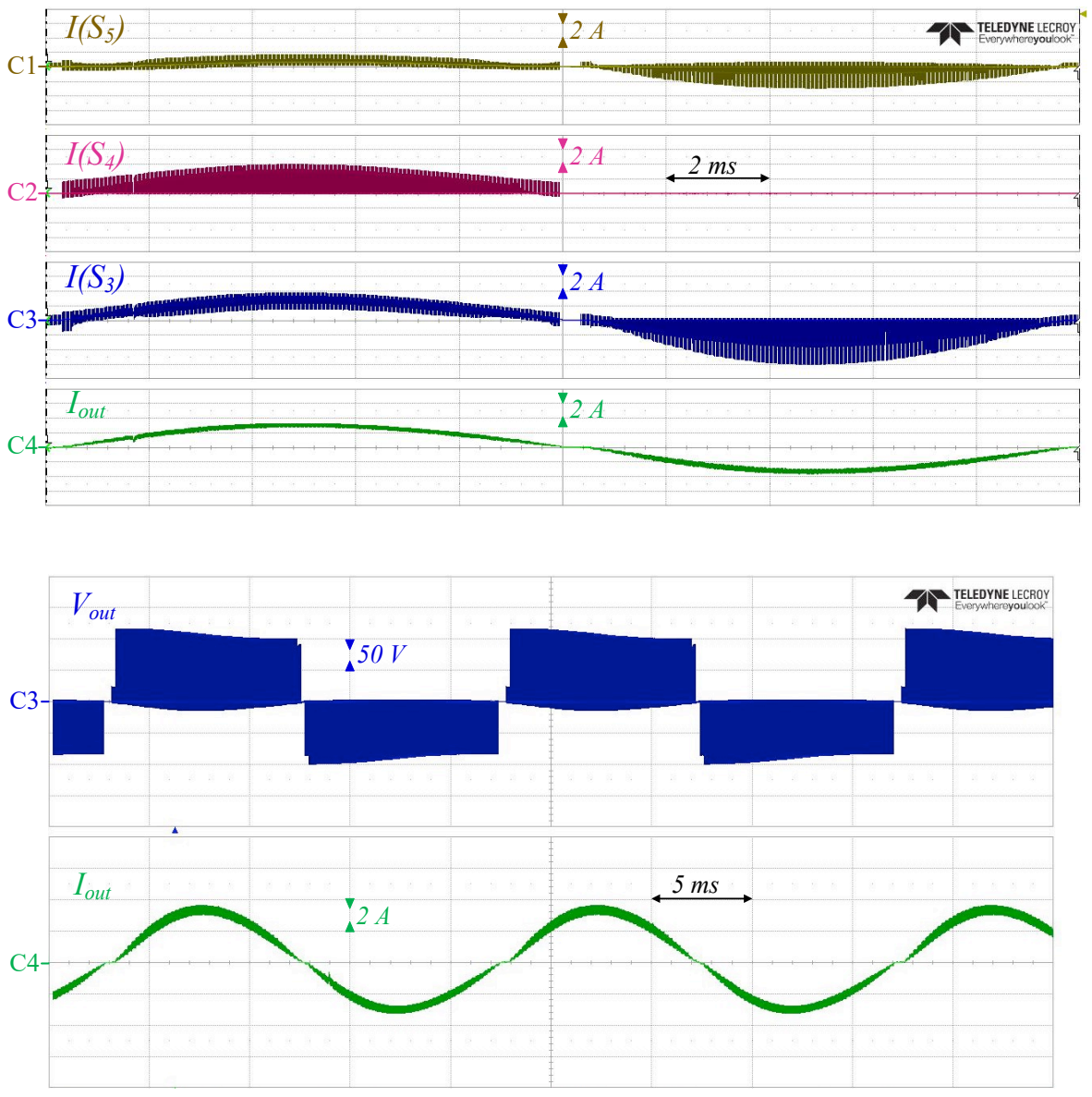

Fig. 12: Implementation of ANPC-FPCM: $V_{in}$=200V $I_{out}$=3A m=0.7.

Preliminary experimental tests of Figs. 11 and 12 display the main electrical quantities of the power converter when the last is operated according to ANPC-SSCM and ANPC-FPCM, respectively. The power converter has been tested under different amplitude modulation indexes measuring the RMS values of the currents flowing through power devices. Such measurements have been compared to that coming from the analytical formulation of Tab. II. The results are graphically represented in Fig. 13, which underline a satisfying agreement between the estimated and measured RMS currents for most of the tests.

Tab. III –Technical specifications of GaN HEMT.

| | |
|---|---|
| $V_{DS}\,(V_{GS} = 0V)$ | *650 V* |
| $I_D$ *@ 100°C* | *10A* |
| $R_{dson}$ | *65mΩ* |
| $V_{GS}$ | *6V/-3V* |

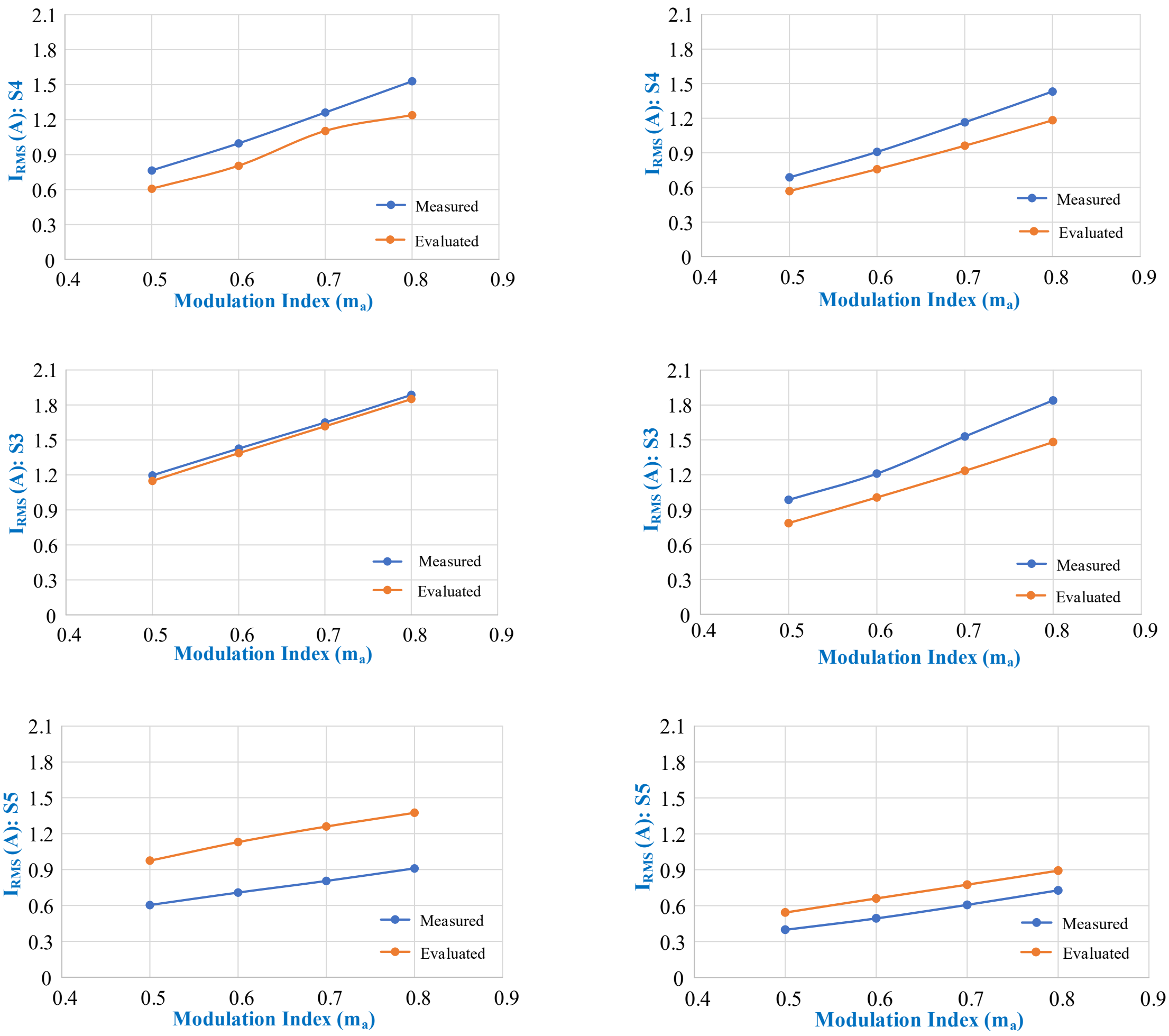

(a) ANPC-SSCM (b) ANPC-FPCM

Fig. 13: RMS Currents determined with the analytical approach and with measurements

# Conclusions

In this paper, a modelling approach to compute the power loss distribution in a three-phase three-level (3L) active neutral point clamped (ANPC) inverter has been investigated, evaluating the difference due to the modulation strategies. Simulations performed with a circuit-based power electronics tool have confirmed the good agreement with results carried out from the analytical model of Tab. II. The effectiveness of the proposed modelling has been also evaluated with some preliminary experimental tests comparing the measured and estimated RMS currents flowing through the power switches. As a main result of the comparison of the four carrier-based modulation techniques, it was confirmed that the modulation with full-path clamping technique allows to get lower total losses due to the effective current sharing strategy.